\documentclass{siamltex}
\usepackage{epsfig}
\usepackage{citesort}

\newtheorem{fact}[theorem]{Fact}

\newcommand{\compmwm}{{\rm Compute-MWM}}
\newcommand{\compcov}{{\rm Compute-Min-Cover}}
\newcommand{\recmwm}{{\rm Recover-Max-Matching}}
\newcommand{\compallcavity}{{\rm Compute-All-Cavity}}
\newcommand{\DDelta}{{\scriptscriptstyle \Delta}}
\newcommand{\overlineG}{G^{\DDelta}}

\newcommand{\cover}{{C}}
\newcommand{\coverUnion}{D}
\newcommand{\overlinecover}{\cover^{\DDelta}}
\newcommand{\calV}{V}
\newcommand{\mwm}{{\rm mwm}}
\newcommand{\mwc}{{\rm mwc}}
\newcommand{\mm}{{\rm mm}}
\newcommand{\fG}{{\phi(G)}}

\renewcommand{\H}{\fG|{\cover_h}}

\begin{document}
\title{A Decomposition Theorem for Maximum Weight Bipartite Matchings\thanks{A 
preliminary version appeared
in the Proceedings of the 7th Annual European Symposium on Algorithms,
1999, pp.\ 439--449.}
}

\author{Ming-Yang Kao\thanks{Department of Computer Science, 
Yale University, New Haven,
CT 06520, USA; kao-ming-yang@cs.yale.edu.
Research supported in part by NSF Grant 9531028.}
\and
Tak-Wah Lam\thanks{Department of Computer Science and Information
Systems, University of Hong Kong, Hong Kong;
\{twlam, wksung, hfting\}@csis.hku.hk.
Research supported in part by Hong Kong RGC Grant HKU-7027/98E.}
\and
Wing-Kin Sung\footnotemark[3]
\and
Hing-Fung Ting\footnotemark[3]}

\maketitle

\begin{abstract}
Let $G$ be a bipartite graph with positive integer weights on the
edges and without isolated nodes.  Let $n$, $N$ and $W$ be the node count,
the largest edge weight
and the total weight of $G$. Let $k(x, y)$ be $\log x / \log (x^2/y)$.
 We present a new decomposition theorem
for maximum weight bipartite matchings and use it to design an
$O(\sqrt{n}W / k(n, W/N))$-time algorithm for
computing a maximum weight matching
of $G$.  This algorithm bridges a long-standing gap between the best
known time complexity of computing a maximum weight matching and that
of computing a maximum cardinality matching.  Given $G$ and a maximum
weight matching of $G$, we can further compute the weight of a maximum
weight matching of $G - \{u\}$ for all nodes $u$ in $O(W)$ time.  
\end{abstract}


\section{Introduction}
Let $G=(X,Y,E)$ be a bipartite graph with positive integer weights on
the edges.  A {\it matching} of $G$ is a subset of node-disjoint edges
of $G$. Let $\mwm(G)$ (respectively, $\mm(G)$) denote the maximum
weight (respectively, cardinality) of any matching of $G$.  A {\it
maximum weight} matching is one whose weight is $\mwm(G)$.  Let $N$ be
the largest weight of any edge.  Let $W$ be the total weight of $G$.
Let $n$ and $m$ be the numbers of nodes and edges of $G$; to avoid
triviality, we maintain $m = \Omega(n)$ throughout the paper.

The problem of finding a maximum weight matching of a given $G$ has a
rich history.  The first known polynomial-time algorithm is the
$O(n^3)$-time Hungarian method \cite{Kuhn55}.  Fredman and Tarjan
\cite{FredmanT87} used Fibonacci heaps to improve the time to $O(n (m
+ n \log n))$.  Gabow \cite{Gabow85} introduced scaling to solve the
problem in $O(n^{3/4} m \log N)$ time by taking advantage of the
integrality of edge weights. Gabow and Tarjan \cite{GaTa} improved the
scaling method to further reduce the time to 
$O(\sqrt{n} m \log (nN))$.  
For the case where the edges all 
have weight 1, i.e., $N=1$ (and $W = m$),
Hopcroft and Karp \cite{HoKa73} gave an $O(\sqrt{n}W)$-time
algorithm, and Feder and Motwani \cite{FederM95} improved
the time complexity to $O(\sqrt{n} W / k(n, m))$,
where $k(x,y) = \log x / \log(x^2 / y)$.
It has remained open 
whether 
the gap between the running times of the Gabow-Tarjan algorithm and 
the latter two algorithms can be closed for
the case where $W=o(m \log (nN))$. 

We resolve this open problem in the affirmative by giving an
$O(\sqrt{n} W / k(n, W/N))$-time algorithm for general $W$.
Note that $W/N = m$ when all the edges have the same weight.
The algorithm does
not use scaling but instead employs a novel decomposition theorem for
weighted bipartite matchings (Theorem~\ref{decom-thm}). 
We also use the theorem to solve the
{\it all-cavity maximum weight matching} problem which, given $G$ and
a maximum weight matching of $G$, asks for $\mwm(G-\{u\})$ for all
nodes $u$ in $G$.  
This problem has applications to tree comparisons \cite{Chung87,kaolst.utree.sjp}.
The case where $N=1$ has been studied by Chung
{\cite{Chung87}}.  Recently, Kao, Lam, Sung, and Ting
\cite{klst-matching.scp} gave an $O(\sqrt{n} m \log N)$-time algorithm
for general $N$.  This paper presents a new algorithm that runs in
$O(W)$ time.

Section \ref{section-mwm-algorithm} presents the decomposition theorem
and uses it to compute the weight of a maximum weight matching.
Section \ref{section-recover-max-weighted-matching} gives an algorithm
to construct a maximum weight matching.  
Section \ref{section-maximum-all-cavity-matching} solves the all-cavity 
matching problem.  

\section{The decomposition theorem}\label{section-mwm-algorithm}
In \S\ref{sec_decomp_thm}, we state the decomposition theorem 
and use
the theorem to design an algorithm to compute
the weight $\mwm(G)$ in 
$O(\sqrt{n}W / k(n, W/N))$ time.  In
\S\ref{sec_decomp_proof}, we prove the decomposition theorem.
In \S\ref{section-recover-max-weighted-matching}, we further construct
a maximum weight matching itself within the same time bound.

\subsection{An algorithm for computing \mathversion{bold}mwm($G$)}
\label{sec_decomp_thm}
Let $\calV(G)$ be the node set of $G$, i.e., $X \cup Y$.
Let $w(u,v)$ denote the weight of an edge $uv \in G$; if $u$ is not
adjacent to $v$, let $w(u,v)=0$. A {\it cover} of $G$ is a function
$C: X \cup Y \rightarrow \{0, 1, 2, \ldots\}$ such that $C(x) + C(y)
\geq w(x,y)$ for all $x \in X$ and $y \in Y$.  
Let $w(C)= \sum_{z \in X \cup Y} C(z)$ be the weight of $C$.
$C$ is a {\it minimum
weight} cover if $w(C)$ is the smallest possible.  
Let $\mwc(G)$ denote the weight of a minimum weight cover of $G$.
A minimum weight cover is a dual of a maximum weight
matching as stated in the next fact.

\begin{fact}[see \cite{BM76}]\label{fact-dual}
Let $C$ be a cover and $M$ be a matching of $G$.  The following
statements are equivalent.
\begin{enumerate}
\item 
$C$ is a minimum weight cover and $M$ is a maximum weight matching of
$G$.
\item 
$\sum_{uv \in M} w(u,v) = \sum_{u \in X \cup Y} C(u)$.
\item 
Every node in $\{ u \mid C(u) > 0 \}$ is matched by some edge in $M$,
and $C(u) + C(v) = w(u, v)$ for all $uv \in M$.
\end{enumerate}
\end{fact}

\begin{figure}
\begin{center}
\epsfig{figure=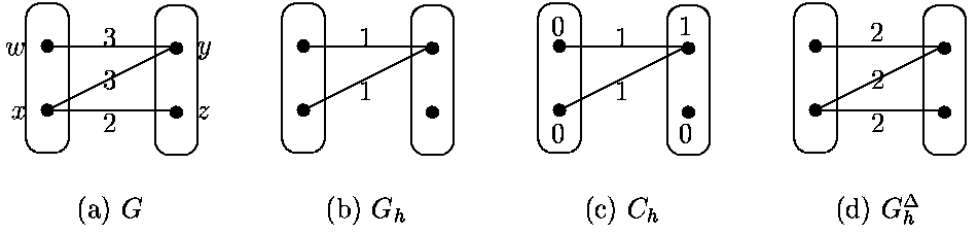, width=\textwidth}
\end{center}
\caption{Consider $h=1$.
$G$ is decomposed into $G_h$ and $\overlineG_h$;
$C_h$ is a minimum weight cover of $G_h$.}
\label{Example1}
\end{figure}

For an integer $h \in [1, N]$, we divide $G$ into two lighter
bipartite graphs $G_h$ and $\overlineG_h$ as follows:
\begin{itemize}
\item $G_h$ is formed by the edges $uv$ of $G$ with $w(u,v) \in
[N-h+1, N]$.  Each edge $uv$ in $G_h$ has weight $w(u,v)-(N-h)$.
For example, $G_1$ is formed by the heaviest edges of $G$, and the
weight of each edge is exactly one.
\item
Let ${\cover}_h$ be a minimum weight cover of $G_h$.  $\overlineG_h$
is formed by the edges $uv$ of $G$ with
$w(u,v)-{\cover}_h(u)-{\cover}_h(v)>0$.  The weight of $uv$ is
$w(u,v) - {\cover}_h(u) - {\cover}_h(v)$.
\end{itemize}
An example is depicted in Figure~\ref{Example1}.
Note that the total weight of $G_h$ and $\overlineG_h$ is at most $W$.

The next theorem is the decomposition theorem. 

\begin{theorem} \label{decom-thm}
$\mwm(G) = \mwm(G_h) + \mwm(\overlineG_h)$; in particular,
$\mwm(G)=\mm(G_1)+\mwm(\overlineG_1)$.
\end{theorem}

\begin{proof}
See \S\ref{sec_decomp_proof}.
\end{proof}

Theorem~\ref{decom-thm} suggests the following recursive algorithm to
compute $\mwm(G)$.

\noindent {\bf Procedure}  \compmwm$(G)$
\begin{enumerate}
\item 
Construct $G_1$ from $G$.
\item 
Compute $\mm(G_1)$ and find a minimum weight cover ${\cover}_1$ of
$G_1$.
\item Construct $\overlineG_1$ from $G$ and $\cover_1$.
\item 
If $\overlineG_1$ is empty, then return $\mm(G_1)$; otherwise, return
$\mm(G_1)+$\compmwm$(\overlineG_1)$.
\end{enumerate}

\begin{theorem} \label{time_for_MWM}
\compmwm$(G)$ finds $\mwm(G)$
in $O(\sqrt{n} W / k(n, W/N))$ time.
\end{theorem}
\begin{proof}
The correctness of \compmwm\ follows from Theorem~\ref{decom-thm}.  Below,
we analyze the running time. We initialize a maximum heap \cite{clr90} in
$O(m)$ time to store the edges of $G$ according to their weights.  Let
$T(n, W, N)$ be the running time of \compmwm\ excluding this
initialization.  Let $L$ be the set of the heaviest edges in $G$.
Then Step~1 takes $O(|L|\log m)$ time.  In Step~2, 
we can compute $\mm(G_1)$ 
in $O(\sqrt{n}|L| / k(n, |L|))$ time~\cite{FederM95}.  
{From} this matching, ${\cover}_1$ can be found in
$O(|L|)$ time \cite{BM76}.  Let $L_1$ be the set of the edges of $G$
adjacent to some node $u$ with ${\cover}_1(u) > 0$; i.e., $L_1$
consists of the edges of $G$ whose weights are reduced in
$\overlineG_1$. 
Let $\ell_1 = |L_1|$.
Step~3 updates every edge of $L_1$ in the heap in
$O(\ell_1 \log m)$ time. As $L \subseteq L_1$, 
Steps~1 to 3 altogether use $O(\sqrt{n}\ell_1/k(n,\ell_1))$ time.
Since the total weight of $\overlineG_1$ is
at most $W - \ell_1$, Step~4 uses at most $T(n,W-\ell_1,N')$ time,
where $N'<N$ is the maximum edge weight of $\overlineG_1$. 
In summary, for some positive integer $\ell_1 \leq W$,
\[ T(n,W,N) = O( \sqrt{n} \ell_1 / k(n, \ell_1)) +
T(n,W - \ell_1,N'),
\]
where $T(n,0,N')=0$.
By recursion, for some positive integers $\ell_1, \ell_2,\ldots,\ell_p$ with
$p \leq N$ 
and $\sum_{1\leq i \leq p} \ell_i=W$,
\begin{eqnarray*}
 T(n,W,N) &=&  O \Bigg(\sqrt{n}\bigg(\frac{\ell_1}{k(n, \ell_1)} +
		   \frac{\ell_2}{k(n, \ell_2)} + \cdots +
		   \frac{\ell_p}{k(n, \ell_p)}\bigg)\Bigg) \\
	&=& O\Bigg(\frac{\sqrt{n}}{\log n}\bigg (\Big(\sum_{1 \leq i \leq p}
				    \ell_i\Big)\log n^2 -
			 	 \sum_{1 \leq i \leq p}
				    \ell_i\log\ell_i \bigg ) \Bigg).
\end{eqnarray*}
Since $x\log x$ 
is convex, by Jensen's Inequality~\cite{HLP88}, 
\[
   \sum_{1 \leq i \leq p} \ell_i\log\ell_i \geq 
		\Big(\sum_{1 \leq i \leq p} \ell_i \Big)
				\log\frac{\sum_{1 \leq i \leq p} \ell_i}{p}
	\geq W \log \frac{W}{N}.
\]
Therefore, 
\begin{eqnarray*}
T(n,W,N) & = &
 O\Bigg(\frac{\sqrt{n}}{\log n} \bigg(W\log n^2 - 
			W\log \frac{W}{N} \bigg)  \Bigg) 
\\
& = &
  O\Bigg(\frac{\sqrt{n}W}{\log n/ \log(n^2/ 
				\mbox{$\textstyle \frac{W}{N}$}\big)}\Bigg) 
         = O\big(\sqrt{n}W/k(n,W/N)\big).
\end{eqnarray*}
\end{proof}

\subsection{Proof of Theorem~\ref{decom-thm}} \label{sec_decomp_proof}
This section proves the statement
that $\mwm(G) = \mwm(G_h) + \mwm(\overlineG_h)$,
where  $\overlineG_h$ is defined according to an arbitrary
minimum weight cover $C_h$ of $G_h$.
By Fact~\ref{fact-dual}, it suffices to prove
$\mwc(G) = w(C_h) + \mwc(\overlineG_h)$.

To show the direction $\mwc(G) \leq w(C_h) + \mwc(\overlineG_h)$,
note that any cover $D$ of $\overlineG_h$ augmented with
$C_h$ gives a cover $C$ of $G$, where
$C(u) = C_h(u) + D(u)$ for each node $u$ of $G$.
Then $C(u)+C(v) \ge w(u,v)$ for all edges $uv$ of $G$.
Thus, $\mwc(G) \le w(C_h) + \mwc(\overlineG_h)$.

To show the direction $w(C_h) + \mwc(\overlineG_h) \le \mwc(C)$,
let $C$ be a minimum weight cover of $G$.  
A node $u$ of $G$ is called {\em bad}\/ if $C(u) < C_h(u)$.
Lemma~\ref{BadCover} below shows
that $G$ must have a minimum weight cover $C$
allowing no bad node.  
Then we can construct a cover $D$ of $\overlineG_h$ as follows.
For each node $u$ of $G$, define $D(u) = C(u) - C_h(u)$,
which must be at least $0$.   $D$ is a cover of $\overlineG_h$
because for any edge $uv$ of $\overlineG_h$,
$D(u) + D(v) = C(u) + C(v) - C_h(u) - C_h(v) 
\ge w(u,v) - C_h(u) - C_h(v)$.
Note that $w(D) = w(C) - w(C_h)$.  Thus,
 $\mwc(\overlineG_h) \le w(C) - w(C_h)$, or
equivalently, $\mwc(\overlineG_h) + w(C_h) \le \mwc(G)$.

The next lemma concludes the proof of Theorem~\ref{decom-thm}.

\begin{lemma}\label{BadCover}
There exists a minimum weight cover of $G$
such that no node of $G$ is bad.
\end{lemma}
\begin{proof}
Suppose, for the sake of contradiction, that every minimum
weight cover allows some bad node.  Then we can obtain
a contradiction by constructing
another minimum weight cover with no bad node.

Let $C$ be a minimum weight cover of $G$ with $u$ as a bad node,
i.e., $C(u) < C_h(u)$.
Recall that $C_h$ is a minimum weight cover of $G_h$.
Consider a maximum weight matching $M$ of $G_h$.
By Fact~\ref{fact-dual}, since $C_h(u) > C(u) \ge 0$,
$u$ is matched by an edge in $M$, say, to a node $v$,
and $C_h(u) + C_h(v) = w(u,v) - (N-h)$.
We call $v$ the {\it mate}\/ of $u$.
Note that $v$ cannot be a bad node; otherwise,
$C(u) + C(v) < w(u,v) - (N-h) \le w(u,v)$ and a contradiction occurs.

Since $C$ is a cover of $G$,
$C(u) + C(v) \ge w(u,v)$.  Thus,
$C(v) \ge w(u,v) - C(u) \ge
  N-h+C_h(u)+C_h(v) - C(u)$.
Define another cover $C'$ of $G$ as follows.
For each bad node defined by $C$, let $v$ be the mate of $u$,
define $C'(u) = C_h(u)$ and
$C'(v) = C(v) - (C_h(u) - C(v))$.
Note that $u$ is not a bad node with respect to $C'$,
and neither is $v$ since $C'(v) \ge N-h+C_h(v) \ge C_h(v)$.
For all other nodes $x$, $C'(x)$ is the same as $C(x)$.
Therefore, if $C'$ is a cover of $G$,
$C'$ allows no bad node.  Also, $w(C') = w(C)$.

It remains to prove that $C'$ is a cover of $G$.
By the definition of $C'$, $C'(v) < C(v)$ 
if and only if $v$ is the mate of a bad node with respect to $C$.
Suppose $C'$ is not a cover of $G$.  Then there
exists an edge $vt$ such that $C'(v) + C'(t) \le w(v,t)$
and $v$ is the mate of a bad node.  Recall that the latter implies
that $C'(v) \ge N-h + C_h(v)$.
In other words,
\[
C'(t) <w(v,t) - C'(v) \le w(v,t) - (N-h) - C_h(v).
\]
We can derive a contradiction as follows.

{\it Case 1:} $w(v,t) \le N-h$. Then $C'(t) < -C_h(v) \le 0$,
which contradicts that $C'(t) \geq C_h(t) \ge 0$.

{\it Case 2:} $w(v,t) > N-h$. Then $G_h$ contains the
edge $vt$ and $C_h(v) + C_h(t) \ge w(v,t) - (N-h)$.
Thus, $C'(t) 
< w(v,t) - (N-h) - C_h(v) \leq C_h(t)$, which
contradicts the fact that $C'$ allows no bad node.

In conclusion, $C'$ is a cover of $G$.  Together with
the fact that $w(C) = w(C')$, we obtain the desired contradiction
that $C'$ is a minimum weight cover of $G$ with no bad node.
Lemma~\ref{BadCover} follows.
\end{proof}

\section{Construct a maximum weight matching}
\label{section-recover-max-weighted-matching}
The algorithm in \S\ref{sec_decomp_thm} only computes the value
of $\mwm(G)$.  To report the edges involved, we show below how to first construct a
minimum weight cover of $G$ in $O(\sqrt{n}W/k(n, W/N))$ time and then use this
cover to construct a maximum weight matching in $O(\sqrt{n}m/k(n,m))$ time.
Thus, the time required to construct a maximum weight matching is
$O(\sqrt{n} W / k(n, W/N))$.

\begin{lemma} \label{lemma-cover-decompose}
Assume that $h, G_h, \cover_h$, and $\overlineG_h$ are defined
as in \S\ref{section-mwm-algorithm}.  
Let $\overlinecover_h$ be any minimum weight
cover of $\overlineG_h$.
If $D$ is a function on $\calV(G)$ such that
for every $u \in \calV(G)$,  
$\coverUnion(u) = \cover_h(u) + \overlinecover_h(u)$,
then
$\coverUnion$ is a minimum weight cover of $G$.
\end{lemma}
\begin{proof}
Consider any edge $uv$ of $G$.  If $uv$ is not
in $\overlineG_h$, then $w(u,v) \leq \cover_h(u) + \cover_h(v)
\leq \coverUnion(u) + \coverUnion(v)$.  Assume that 
$uv$ is in $\overlineG_h$.
Note that its weight 
in $\overlineG_h$ is $w(u, v) - \cover_h(u) - \cover_h(v)$.
Since $\overlinecover_h$ is a  cover,
$\overlinecover_h(u) + \overlinecover_h(v) \geq
w(u, v) - \cover_h(u) - \cover_h(v)$.
Thus, $\coverUnion(u) + \coverUnion(v) = 
\cover_h(u) + \overlinecover_h(u) +
\cover_h(v) + \overlinecover_h(v)
\geq w(u, v)$. It follows that
$\coverUnion$ is a  cover of $G$.
To show that $\coverUnion$ is a minimum weight one, 
we observe that
\[ \begin{array}{llll}
\sum_{u \in \calV(G)} \coverUnion(u) & = &
  \sum_{u \in \calV(G)} \cover_h(u) + \overlinecover_h(u) \\
& = & \sum_{u \in \calV(G)} \cover_h(u) +
  \sum_{u \in \calV(G)} \overlinecover_h(u) \\
& = & \mwm(G_h) + \mwm(\overlineG_h) & \mbox{by Fact~\ref{fact-dual}}\\
& = & \mwm(G). & \mbox{by Theorem~\ref{decom-thm}}
\end{array} \]
By Fact~\ref{fact-dual}, $D$ is minimum.
\end{proof}

By Lemma~\ref{lemma-cover-decompose}, a minimum weight cover of
$G$ can be computed using a recursive procedure similar to \compmwm\ 
as follows.

\newcommand{\cocover}{\cover^\Delta}

\noindent {\bf Procedure} \compcov$(G)$
\begin{enumerate}
\item Construct $G_1$ from $G$.
\item Find a minimum weight cover ${\cover}_1$ of $G_1$.
\item Construct $\overlineG_1$ from $G$ and $C_1$.
\item If $\overlineG_1$ is empty,
  then return $C_1$; otherwise, let
$\cocover_1$ = \compcov$(\overlineG_1)$ and
return $D$ where for all nodes $u$ in $G$,
$D(u) = \cover_1(u) + \cocover_1(u)$.
\end{enumerate}

\begin{theorem} \label{thm-min-cover}
\compcov$(G)$ correctly computes a minimum weight cover of $G$ in
$O(\sqrt{n}W/k(n, W/N))$ time.
\end{theorem}
\begin{proof}
The correctness of \compcov$(G)$ follows
from Lemma~\ref{lemma-cover-decompose}.
For the time complexity,
the analysis is similar to that of Theorem~\ref{time_for_MWM}.
\end{proof}

Now, we show how to recover a maximum weight matching of $G$
from a minimum weight cover $\coverUnion$ of $G$.

\noindent {\bf Procedure} \recmwm$(G, \coverUnion)$
\begin{enumerate}
\item 
Let $H$ be the subgraph of $G$ that contains all edges $uv$
with $w(u,v) = \coverUnion(u) + \coverUnion(v)$.
\item Make two copies of $H$. Call them $H^a$ and $H^b$.
For each node $u$ of $H$, let $u^a$ and $u^b$ denote the
corresponding nodes in $H^a$ and $H^b$, respectively.
\item Union $H^a$ and $H^b$ to form $H^{ab}$, and add to $H^{ab}$ the set
of edges $\{ u^au^b \mid u \in \calV(H), ~{\coverUnion}(u) = 0 \}$.
\item
Find a maximum cardinality matching $K$ of $H^{ab}$
and return the matching $K^a = \{uv \mid u^av^a \in K\}$.
\end{enumerate}

\begin{theorem} \label{thm-extract-matching}
\recmwm$(G,D)$ correctly computes a maximum
weight matching of $G$ in $O(\sqrt{n}m/k(n,m))$ time.
\end{theorem}
\begin{proof}
The running time of \recmwm$(G,D)$
is dominated by the construction of $K$.
Since $H^{ab}$ has at most $2n$ nodes and at most 
$3m$ edges, $K$ can be constructed in
$O(\sqrt{n}m/k(n,m))$ time using 
Feder-Motwani algorithm \cite{FederM95}.

It remains to show that $K^a$ is a maximum weight matching
of $G$.
First, we argue that $H^{ab}$ has a perfect matching.
Let $M$ be a maximum weight matching of $G$.
By Fact~\ref{fact-dual},
$\coverUnion(u) + \coverUnion(v) = w(u, v)$
for every edge $uv \in M$.
Therefore, $M$ is also a matching of $H$.
Let $U$ be the set of nodes in $H$ unmatched by $M$.
By Fact~\ref{fact-dual},
$\coverUnion(u) = 0$ for all $u \in U$.
Let $Q$ be $\{ u^au^b \mid u \in U \}$.
Let $M^a = \{ u^av^a \mid uv \in M \}$
and $M^b = \{ u^bv^b \mid uv \in M \}$.
Note that $Q \cup M^a \cup M^b$ forms a matching in $H^{ab}$ and
every node in $H^{ab}$ is matched by
either $Q$, $M^a$ or $M^b$.
Thus, $H^{ab}$ has a perfect matching.

Since $K$ is a maximum cardinality matching of $H^{ab}$,
$K$ must be a perfect matching.
For every node $u$ with $\coverUnion(u) > 0$,
$u^a$ must be matched by $K$.
Since there is no edge between $u^a$ and any $x^b$ in $H^{ab}$,
there exists some $v^a$ with $u^av^a \in K$.
Thus,
every node $u$ with $\coverUnion(u) > 0$ must be
matched by some edge in $K^a$.
Therefore, $\sum_{uv \in K^a} w(u, v) =
\sum_{u \in X \cup Y, \coverUnion(u)>0} \coverUnion(u) =
\sum_{u \in X \cup Y} \coverUnion(u) =
\mwm(G)$, and $K^a$ is a maximum weight matching of $G$.
\end{proof}

\section{All-cavity maximum weight matchings}
\label{section-maximum-all-cavity-matching}
In \S\ref{sec_unfold}, we introduce the notion of an {\em unfolded graph}.
In \S\ref{sec_cavity_alg}, we use this notion
to design an algorithm which,
given a weighted bipartite graph $G$ and
a maximum weight matching of 
$G$, computes $\mwm(G-\{u\})$ for all nodes $u$ in $G$ using $O(W)$ time.

\subsection{Unfolded graphs} \label{sec_unfold}
\begin{figure}
\begin{center}
\epsfig{figure=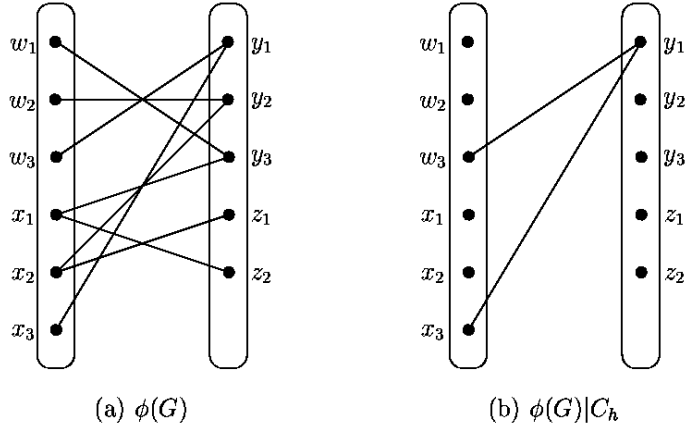, width=0.7\textwidth}
\end{center}
\caption{(a) The unfolded graph $\fG$ of
the bipartite graph given in Figure~\ref{Example1}(a).
(b) With respect to the cover $C_h$
defined in Figure~\ref{Example1}(c),
the node $y_1$ in $\fG$ is the only node satisfying
the condition that $1 \leq C_h(y)$.  Thus,
$\H$ comprises only the edges incident to $y_1$.}
\label{Example2}
\end{figure}

The {\em unfolded graph}\/ $\fG$ of $G$
is defined as follows.  
\begin{itemize}
\item For each node $u$ of $G$, $\fG$ has $\alpha$ copies of $u$,
denoted as $u^1, u^2, \ldots, u^\alpha$, where $\alpha$ is the weight
of the heaviest edge incident to $u$.
\item For each edge $uv$ of $G$, $\fG$ has the edges
$u^1v^\beta,u^2v^{\beta-1},\ldots,u^{\beta}v^1$, where
$\beta=w(u,v)$.
\end{itemize}
See Figure~\ref{Example2}(a) for an example.
Let $M$ be a matching of $G$.  Consider $M$ as a weighted bipartite graph;
then, by definition, 
\(\phi(M)=\bigcup_{uv \in M} \{
u^1v^{\beta}, 
\cdots,u^{\beta}v^1 \mid \beta = w(u, v)\}\) 
is a matching of $\fG$.  
The number of edges in $\phi(M)$
is equal to the total weight of the edges in $M$,
i.e., $|\phi(M)| = \sum_{uv \in M}w(u,v)$.
The next lemma  relates $G$ and $\fG$.  

\begin{lemma}\label{lemma_M_fi} 
Assume that $M$ is a maximum weight matching of $G$.  
\begin{enumerate}
\item\label{lemma_f(M)_1}
	$\mwm(G)=\mm(\fG)$.
\item\label{corollary_f(M)_1}
The set $\phi(M)$ is
a maximum cardinality matching of $\phi(G)$.
\end{enumerate}
\end{lemma}
\begin{proof} Statement \ref{corollary_f(M)_1} follows from Statement
\ref{lemma_f(M)_1}. Statement \ref{lemma_f(M)_1} is proved as follows.
Since $M$ is a maximum weight matching of $G$,  
$\mwm(G) = \sum_{uv \in M} w(u,v) = |\phi(M)| \le \mm(\fG)$.
By Fact~\ref{fact-dual}, 
$\mwm(G) \ge \mm(\fG)$ if and only if $\mwc(G) \ge \mwc(\fG)$.
We prove the latter as follows.
Given a minimum weight cover $C$ of $G$,
we can obtain a cover $C'$ of $\phi(G)$ as follows.
For any node $u$ of $G$,
$C'(u^i) = 1$ if $C(u) >0$ and $i \le C(u)$;
otherwise, $C'(u^i) = 0$.
Note that $w(C') = w(C) = \mwc(G)$.  
Therefore, $\mwc(G) \ge \mwc(\fG)$ and
$\mwm(G) \ge \mm(\fG)$.
\end{proof}

\subsection{An algorithm for all-cavity maximum weight matchings}\label{sec_cavity_alg}
Let $M$ be a given maximum weight matching of $G$.

By Lemma~\ref{lemma_M_fi}(\ref{corollary_f(M)_1}),
$\phi(M)$ is a maximum cardinality matching of $\fG$.
In light of this maximality, we say 
that a path in
$\fG$ is {\it alternating} for $\phi(M)$ if (1) its edges alternate between
being in $\phi(M)$ and being not in $\phi(M)$ and (2) 
in the case the first (respectively, last) node is matched
by $\phi(M)$, the path contains the matched edge of $u$
as the first (respectively, last) edge.
The length of an alternating path is its number of edges.
An alternating path may have zero length; in this case,
the path contains exactly one unmatched node.
An alternating path $P$ can modify $\phi(M)$ to another
matching, i.e., $(\phi(M) \cup P) - (\phi(M) \cap P)$.
If $P$ is of even length, the resulting matching has the
same size as $\phi(M)$.
If $P$ is of odd length, $P$ modifies $M$
to a strictly smaller or bigger matching; yet the latter is impossible
because $\phi(M)$ is maximum.
Intuitively, we would like to maximize the size of
the resultant matching and even-length alternating
paths are preferred.

Our new algorithm for computing $\mwm(G-\{u\})$ 
is based on the observation that $\mwm(G-\{u\})$ can be determined 
by detecting the smallest $i$ such that $u^i$ has an
even-length alternating path for $\phi(M)$.  Details are as follows.

Definition.
For each $u^i$ in $\phi(G)$, let $\rho(u^i) = 0$ if there is an
even-length alternating path for $\phi(M)$ starting from $u^i$;
otherwise, let $\rho(u^i) = 1$.

The following lemma states a monotone property
of $\rho(u^i)$ over different $i$'s.

\begin{lemma} \label{disjoin-path}
Consider any node $u$ in $G$.
Let $u^1, u^2, \ldots, u^{\beta}$ be its corresponding nodes in $\fG$.
If $\rho(u^i) = 0$, then 
$\rho(u^j) = 0$ for all $j \in [i,\beta]$.
Furthermore, there exist  $\beta - i + 1$ node-disjoint
even-length alternating paths $P_i, P_{i+1},\ldots P_{\beta}$ for
$\phi(M)$, where each $P_j$ starts from $u^j$.
\end{lemma}
\begin{proof} 
As $\rho(u^i) = 0$,
let $P_i = u_0^{a_0}, v_0^{b_0},$
$u_1^{a_1}, v_1^{b_1}, \ldots,
u_{p-1}^{a_{p-1}}, v_{p-1}^{b_{p-1}}, u_p^{a_p}$ 
be a shortest even-length alternating path
for $\phi(M)$ where $u_0^{a_0} = u^{i}$.

Based on $P_i$, we can construct
an even-length alternating path $P_{i+1}$ for $\phi(M)$
starting from $u^{i+1}$ as follows.
If $u^{i+1}$  is not matched by $\phi(M)$, $P_{i+1}$ is simply
a path of zero length. From now on, we
assume that $u^{i+1}$ is matched by $\phi(M)$.
As $P$ is of even length, $u_p^{a_p}$ is not matched by $\phi(M)$.
Then, by the definition of $\phi(M)$, 
$u_p^{a_p+1}$ is also not matched by $\phi(M)$.
Let $h$ be the smallest integer in $[1,p]$
such that $u_h^{a_h+1}$ is not matched by $\phi(M)$.
Notice that, for all $\ell < h$,
 $u_\ell^{a_\ell+1}$ is matched to $v_\ell^{b_\ell-1}$;
furthermore, $\phi(G)$ contains an edge
between $v_\ell^{b_\ell-1}$ and $u_{\ell+1}^{a_{\ell+1}+1}$.
Thus, $P_{i+1} = u^{i+1}, v_0^{b_0-1},
u_1^{a_1+1}, v_1^{b_1-1},$ $\cdots, u_h^{a_h+1}$
is an even-length alternating path for $\phi(M)$.
Similarly, 
for $j = i+2, \cdots, \beta$, we can use
$P_i$ to define an
even-length alternating path $P_j$ for $\phi(M)$
starting from $u^j$.
By construction,  $P_i, P_{i+1}, \cdots P_{\beta}$
are node-disjoint.
\end{proof}

The next lemma is the basis of our cavity matching algorithm.
It shows that given $\mwm(G)$ (i.e., the
weight of $M$),
we can compute $\mwm(G-\{u\})$ 
from the values $\rho(u^i)$, and all the $\rho(u^i)$'s
can be found in $O(W)$ time.

\begin{lemma} \label{lemma-weighted-cavity}
\begin{enumerate}
\item\label{stmt-cavity} $\sum_{1 \leq i \leq \beta} \rho(u^i) =
	\mwm(G)-\mwm(G-\{u\})$. 
\item\label{stmt-find-even}
For all $u^i \in \fG$, $\rho(u^i)$ can be computed
in $O(W)$ time in total.
\end{enumerate}
\end{lemma}
\begin{proof}
The two statements are proved as follows.

Statement~\ref{stmt-cavity}.
Let $k$ be the largest integer such that
$\rho(u^k) = 1$.  By Lemma~\ref{disjoin-path},
$\rho(u^i)=1$ for all $1\leq i \leq k$, and
$0$ otherwise. Note that if $\rho(u^i)=1$, 
$u^i$ must be matched by $\phi(M)$.
Thus, $\sum_{1 \leq i \leq \beta}\rho(u^i) = k$.
Below, we prove the following two equalities:
\begin{itemize}
\item[(1)] $\mm(\fG-\{u^1,\ldots, u^k\}) = \mm(\fG) -k$.
\item[(2)] 
  $\mm(\fG-\{u^1,\ldots, u^{\beta}\}) = 
     \mm(\fG-\{u^1, \ldots, u^k\})$.
\end{itemize}
Then, by Lemma~\ref{lemma_M_fi}, $\mwm(G) = \mm(\fG)$ and
  $\mwm(G-\{u\}) = \mm(\fG - \{u^1, \ldots, u^{\beta}\})$.
Thus,  $\mwm(G)-\mwm(G-\{u\})=k$ and Statement~\ref{stmt-cavity} follows.

To show Equality (1),
let $H$ be the set of edges of $\phi(M)$ incident to
$u^i$ with $1 \leq i \leq k$.  Let $M' = \phi(M) - H$.
Then, $|M'|=|\phi(M)|-k$.
We claim that $M'$ is a maximum cardinality matching of
$\phi(G)-\{u^1,...,u^k\}$.
Hence, $\mwm(\phi(G)-\{u^1,...,u^k\}) =
|\phi(M)| -k$, and Equality (1) follows.
We prove the claim by contradiction.
Suppose $M'$ is not a maximum cardinality matching of
$\fG - \{u^1, \ldots, u^k \}$.
Then, there exists an alternating path $P$ that 
can modify $M'$ to a larger matching 
of $\fG - \{u^1, \ldots, u^k \}$ \cite{Galil86,Gerards95};
in particular, the length of $P$ must be odd and
both of its endpoints are not matched by $M'$.
$P$ must start from some node
$v^j$ with $u^iv^j \in \phi(M)$ and $i < k$;
otherwise, $P$ is alternating for $\phi(M)$ in $G$
and $\phi(M)$ cannot be a maximum cardinality matching of $\fG$.
Let $Q$ be a path formed by joining $u^iv^j$ with
$P$. $Q$ is an even-length alternating path for $\phi(M)$
starting from $u^i$ in $\fG$.
This contradicts the fact that there is no even-length alternating
path for $\phi(M)$ starting from $u^i$ for $i < k$.

To show Equality (2),  we first note that 
  $\mm(\fG-\{u^1,\ldots, u^{\beta}\}) \leq 
     \mm(\fG-\{u^1, \ldots, u^k\})$.
It remains to prove the other direction.
By Lemma~\ref{disjoin-path},
we can find
$\beta - k$ node-disjoint even-length alternating 
paths $P_{k+1}, \ldots, P_{\beta}$ for $\phi(M)$, which
start from $u^{k+1}, \cdots, u^\beta$.
$P_j$ starts at $u^j$.
Let $M'' = (\phi(M) \cup (P_{j+1} \cup \cdots \cup
P_\beta)) - (\phi(M) \cap (P_{j+1} \cup \cdots \cup P_\beta))$.
Note that $|M''| = |\phi(M)|$ and
there are no edges in $M''$ incident to any 
of $u^{k+1}, \cdots, u^{\beta}$.
$M''$ is a matching of $\fG-\{u^{k+1}, \cdots, u^{\beta}\}$ and
 $M''-H$ of $\fG-\{u^1,\ldots, u^{\beta}\}$.
$|M''-H| \geq |M''|-k = |\phi(M)|-k$.  Since
$\mm(\fG-\{u^1, \ldots, u^k\}) = |\phi(M)| -k$ by Equality (1),
it follows that
$\mm(\fG-\{u^1,\ldots,u^{\beta}\}) \ge |M''-H| \ge
\mm(\fG-\{u^1,\ldots,u^k\})$.  Therefore, Equality (2) holds.

Statement~\ref{stmt-find-even}.
We want to determine 
whether $\rho(u^i) = 0$ for all nodes $u^i \in \phi(G)$ in $O(W)$ time.
By definition, $\rho(u^i) = 0$ if and only if
there is an even-length 
alternating path for $\phi(M)$ starting from $u^i$.
Let us partition the nodes of $\phi(G)$ into two parts:
 $\phi(X) = \{u^i \in \phi(G) \mid u \in X\}$ and
 $\phi(Y) = \{u^i \in \phi(G) \mid u \in Y\}$.
Below, we give the details of computing
$\rho(u^i)$ for all $u^i \in \phi(X)$.
The case where $u^i \in \phi(Y)$ is symmetric.

Let $D$ be a directed graph over the node set $\phi(X)$.
$D$ contains an edge $u^iv^j$ if there exists a node $w^k \in \phi(Y)$ 
such that $u^iw^k \in \phi(G) - \phi(M)$ and $w^kv^j \in \phi(M)$.
Consider any node $v^j$ of $D$ that is unmatched by $\phi(M)$.
A directed path in $D$ from $v^j$ to a node $u^i$
corresponds to a path in $\phi(G)$, which is indeed
an even-length alternating path for $\phi(M)$ starting from $u^i$.  
Therefore,
for any $u^i \in \phi(X)$, $\rho(u^i) = 0$ if and only if $u^i$ is
reachable from some node in $D$ that is unmatched by $\phi(M)$.
We can identify all such $u^i$ by using a depth-first search 
on $D$ starting with all the nodes unmatched by $M$.
The time required is $O(|D|)$.  As $|D| \leq |\phi(G)| = W$,
the lemma follows.
\end{proof}

The following procedure computes $\mwm(G-\{u\})$ for all nodes $u$ of $G$.
Let $M$ be a maximum weight matching of $G$.

\noindent {\bf Procedure}  \compallcavity$(G,M)$
\begin{enumerate}
\item Construct $\fG$ and $\phi(M)$.
\item For every $j \in [0,n/2]$, determine  $A_j$ from $\phi(M)$.
\item For every node $u^i$ of $\fG$, if $u^i \in \bigcup_j A_j$ then
	$\rho(u^i) = 0$; otherwise $\rho(u^i) = 1$. 
\item For every node $u$ of $G$, compute $\mwm(G-\{u\}) = \mwm(G)-
		\sum_{1 \leq i \leq \beta}\rho(u^i)$ where
	$u^1$, $u^2$,
	$\ldots$, $u^\beta$ are the nodes corresponding to $u$ in $\fG$.
\end{enumerate}
 
\begin{theorem} \label{time_for_all-cavity}
\compallcavity$(G,M)$ correctly computes $\mwm(G - \{u\})$ for all
$u$ of $G$ in $O(W)$ time.
\end{theorem}
\begin{proof}
Follows from Lemma~\ref{lemma-weighted-cavity}
\end{proof}

\section*{Acknowledgments}
The authors wish to thank the anonymous referee for extremely helpful
comments, which significantly improved the presentation of the paper.
In particular, Theorem~\ref{decom-thm} was originally proved using
unfolded graphs (see the conference version of this paper
\cite{klst.mat.scp}); the new proof is based on a suggestion by the
referee.


\end{document}